\documentclass[aps,pre,floats,twocolumn,showpacs,reprint, superscriptaddress]{revtex4-1}

\usepackage{graphicx,color,xcolor}
\usepackage{dcolumn}
\usepackage{bm}

\def\Barabasi{Barab{\'a}si}
\newcommand{\revisiontwo}[1]{#1}
\newcommand{\lay}[1]{^{[#1]}}

\newcommand{\rom}[1]{\uppercase\expandafter{\romannumeral#1}}
\newcommand{\citsymbol}[1][{}]{N^{\rm cit}_{#1}}

\newcommand{\avg}[1]{\langle #1 \rangle}

\begin{document}


\title{The advantages of interdisciplinarity in modern science}

\author{Moreno Bonaventura}
\affiliation{%
 School of Mathematical Sciences, Queen Mary University of London, Mile End Road, E1 4NS, London (UK)
}
\author{Vito Latora}
\affiliation{%
 School of Mathematical Sciences, Queen Mary University of London, Mile End Road, E1 4NS, London (UK)
}
\affiliation{Dipartimento di Fisica e Astronomia, Universit\`a di 
Catania and INFN, 95123 Catania, Italy}  

\author{Vincenzo Nicosia}
\affiliation{%
 School of Mathematical Sciences, Queen Mary University of London, Mile End Road, E1 4NS, London (UK)
}
\author{Pietro Panzarasa}
\affiliation{
 School of Business and Management, Queen Mary University of London, Mile End Road, E1 4 NS, London (UK)
}

\date{\today}

\begin{abstract}

As the increasing complexity of large-scale research requires the
combined efforts of scientists with expertise in different fields, the
advantages and costs of interdisciplinary scholarship have taken center 
stage in current debates on scientific production. Here we
conduct a comparative assessment of the scientific success of
specialized and interdisciplinary researchers in modern
science. Drawing on comprehensive data sets on scientific
production, we propose a two-pronged approach to
interdisciplinarity. For each scientist, we distinguish between
background interdisciplinarity, rooted in knowledge accumulated over
time, and social interdisciplinarity, stemming from exposure to
collaborators' knowledge. We find that, while abandoning
specialization in favor of moderate degrees of background
interdisciplinarity deteriorates performance, very interdisciplinary
scientists outperform specialized ones, at all career stages.
Moreover, successful scientists tend to intensify the heterogeneity of
collaborators and to match the diversity of their network with the
diversity of their background. Collaboration sustains performance by
facilitating knowledge diffusion, acquisition and creation. Successful
scientists tend to absorb a larger fraction of their collaborators'
knowledge, and at a faster pace, than less successful
ones. Collaboration also provides successful scientists with
opportunities for the cross-fertilization of ideas and the synergistic
creation of new knowledge. These results can inspire scientists to
shape successful careers, research institutions to develop effective
recruitment policies, and funding agencies to award grants of enhanced
impact.

\end{abstract}

\maketitle


The debate on the comparative benefits of specialized and
interdisciplinary scholarship boasts a longstanding tradition
stretching back several centuries, and still remains largely
controversial and unresolved \cite{Aldrich,Jacobs,Nature,Noorden2015,Whitfield}. The idea that knowledge can be organized into distinct and self-contained disciplinary fields can be traced as
far back as ancient Greek philosophy, and arguments in favor of a
hierarchical structure of knowledge proved remarkably resilient over
time \cite{Jacobs}. Equally, the quest for tighter connections among
disciplines is not new. By the late Middle Ages, the growth of universities was marked by an emphasis on the
universality of knowledge transcending disciplinary boundaries. A new
fertile terrain for interdisciplinary scholarship was then found in
the Renaissance movement. Indeed it is widely accepted that the finest minds of the Western intellectual tradition, from Leonardo da Vinci and Pico della Mirandola to Copernicus, were characterized by an extraordinary
ability to master the breadth and depth of the disciplinary
landscape. Calls for `the totality of sciences and arts"
\cite[p.19]{Vico} continued until modern times, and co-existed with
a new focus that the Enlightenment enterprise placed on intellectual classification. 

In more recent times, the debate on specialization and
interdisciplinarity has been reshaped by the institutional changes
spurred by the growth of modern universities and by the mounting
pressure faced by the scientific community in connection with
competitive funding, research assessment, and publication in
top-ranked academic journals \cite{Jacobs}. On the one hand, the rapid
proliferation of new specialties has raised the costs that scientists
bear for working outside their usual fields \cite{Leahley}. On the other, the
increasing complexity of large-scale projects has prompted an increase
in teamwork combining the efforts of multiple scientists with
expertise in different fields \cite{shena,Wuchty2007}. Interestingly,
the case for interdisciplinarity has recently been embraced by the
same academic institutions that have contributed to the consolidation
of specialized knowledge and intellectual hierarchies. The addition of
``interdisciplinarity'' as a special category of funding at prominent
research institutes and as a submission category in highly regarded
academic journals is a clear indication of recent cultural
orientations toward a more integrated scientific scholarship
\cite{Aldrich}.

In this article we contribute to this debate by studying how scientific performance varies as the scientist abandons specialization in favor of a more inclusive and integrative approach to research or vice versa. Our analysis casts light on the mechanisms,
the career paths and the research strategies that nurture and sustain scientific success over time \cite{Chakraborty}. To this end, we
propose a dual-faceted perspective on interdisciplinarity that blends
individual learning with social exposure. To undertake research at the
interface among disciplines, scientists can either acquire new
expertise on their own, through ``in-breadth'' learning and training,
or seek collaborators with the necessary knowledge and experience. In
turn, exposure to collaborators' knowledge may induce absorption
or the synergistic creation of knowledge. Here we examine the
implications of these research strategies for scientific success using
two large-scale network data sets on scientific collaborations and
citations, and we uncover the comparative benefits of
interdisciplinary and specialized careers.

\begin{figure*}[ht]
   \includegraphics[width=6.1in]{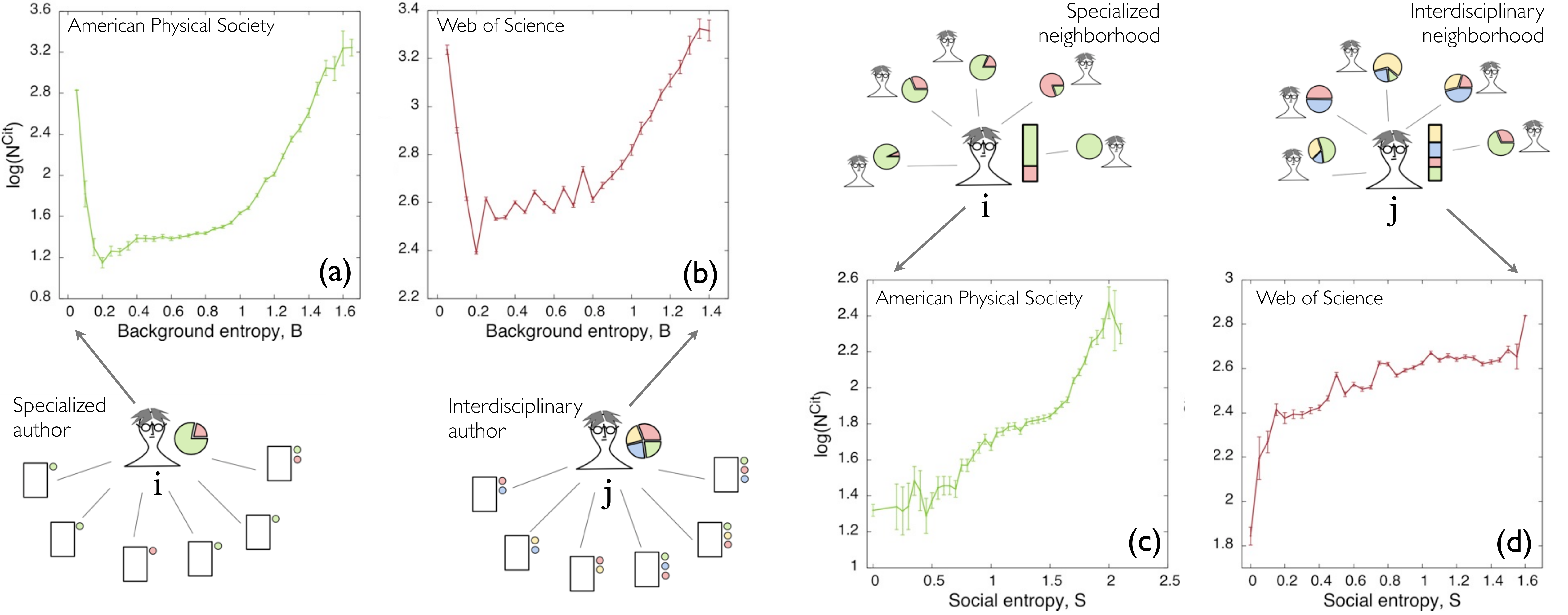}
   \caption{\textbf{Interdisciplinarity and success.}   
       Scientific success depends on authors' background and social
       entropy, both in physics (panels a,c) and in the natural
       sciences (panels b,d). The background entropy $B_i$ of an
       author $i$ quantifies the heterogeneity of the research topics
       with which $i$'s scientific production is concerned, whilst the
       social entropy $S_i$ reflects the variety of the knowledge to
       which author $i$ is exposed through the collaborators. The
       vignettes below panels a,b show that, even if two authors $i$
       and $j$ have published the same number of articles, the of 
       personal codes $PC_i$ and $PC_j$ (represented by the
       pie-charts) might differ: one author might focus just on a few
       scientific subjects (author $i$, with a small value of $B_i$),
       and the other might be interested in several different subjects
       (author $j$, with a larger value of $B_j$). Nevertheless, both
       authors, one specialized and the other interdisciplinary, can
       have a comparable level of scientific success, as indicated in
       panels a,b by the U-shaped line, which represents the average
       number of citations accumulated in his or her career by an
       author with a given value of background entropy. Similarly, as
       shown in the vignettes above panels c,d, the number of
       collaborators is not a predictor of the variety of knowledge to
       which authors are exposed through their networks (represented
       by the vertical color bars). Panels c,d indicate the average
       number of citations obtained by an author with a given value of
       social entropy. On average, authors whose collaborators are
       focused just on a few subjects (and with a small value of $S$)
       tend to be outperformed by authors with a more heterogeneous
       network (and with a high value of $S$). Error bars represent
       the standard error of the mean.}
\label{fig:fig1}
\end{figure*}

From the vantage point of
individual scientists, we show that \textit{diversity trumps focus}. In so doing, we do not argue that diverse groups outperform focused individuals \cite{Singh}, nor that there are economies of scale associated with group production \cite{Wuchty2007}. \revisiontwo{Rather, we suggest that individual scientists can reap the rewards of intellectual diversity by \textit{forsaking ``in-depth'' learning for a suitable admixture of ``in-breadth'' learning and social networking}.}

\section{Results}
We study the association of interdisciplinarity with
  scientific success at two levels of analysis. To this end, we
  examine co-authorship networks, citations and research fields
  drawing on articles included in two databases : (i) the American Physical Society (APS)
  (micro level) \cite{apsdata}; and (ii) the Web of Science (WOS)
  (macro level) \cite{webofscience}. First, from the APS data set we
  extracted $380,913$ articles published after 1980 and authored or
  co-authored by $136,871$ scientists whose careers started after
  1980. Each article is associated with up to four codes included in
  the Physics and Astronomy Classification Scheme (PACS). We obtained
  $1,154$ distinct PACS codes identifying the research areas to which
  each article belongs, and used these codes to measure
  interdisciplinarity at a micro level (i.e., in physics)
  ~\cite{Pan2012,Sinatra2015}. Second, from the WOS data set we
  extracted $1,125,729$ articles published by $1,532,673$ scientists
  between 1945 and 2014 in the top five journals with the highest
  impact factor in $50$ different research categories concerned with
  the natural sciences, including biology, chemistry, computer
  science, mathematics, and physics. The breadth of scientific production embodied in
  the WOS data set enables us to conduct our study at a higher level,
  across disciplinary boundaries. Our analysis relies on a conservative method for disambiguating authors' names, based on institutional affiliations, collaboration network, and citation network \cite{Deville2014}.

\subsection{Quantifying interdisciplinarity}
We use the PACS codes and research categories
  (respectively, for the APS and WOS data sets) associated with the
  articles of an author to identify the author's research interests
  and expertise. To measure author $i$'s interdisciplinarity
\cite{Leahley,Chakraborty,Porter,Sayama,Schummer}, we construct the list
 $\bm{PC}_i$ of \textit{personal
  codes or categories}, defined as the PACS codes or research
categories extracted from all the articles published by $i$ during
$i$'s scientific career. The list $\bm{PC}_i$ thus reflects the disciplinary areas to which author
$i$ has contributed, and can be used as a proxy for $i$'s (cumulative)
background knowledge \cite{Whitfield, Schummer}. We measure author
$i$'s \textit{background interdisciplinarity} through the
\textit{background entropy} defined as the Shannon entropy of the list
$\bm{PC}_i$ of the author's personal PACS codes or research categories
~\cite{shannon}:
\begin{equation}
  B_i = - \sum_{\alpha} p\lay{\alpha}_i \log ( p\lay{\alpha}_i ), 
\label{eq:background_entropy}
\end{equation}
where $p\lay{\alpha}_i =
\frac{n\lay{\alpha}_i}{\sum_{\beta}n\lay{\beta}_i}$, and
$n_i\lay{\alpha}$ is the number of times the PACS code or research
category $\alpha$ is found in $\bm{PC}_i$ (i.e., the number of
articles authored by $i$ that belong to $\alpha$). Similar
entropy-based measures have been used for quantifying the
heterogeneity of the citations made by an
article~\cite{Leahley,Porter,Lariviere2010}. In general, authors with a more heterogeneous
background are characterized by higher values of $B$, whilst smaller
values are typically associated with authors whose research is focused
on a small number of scientific sub-fields or categories.

By forging collaborations, scientists are exposed to various sources
of knowledge, which may not be entirely coextensive with their own
personal background, and on which they can rely to widen the
scientific horizons of their research. To assess scientists' exposure
to their collaborators' knowledge, we propose a measure that is meant
to directly capture the social roots of interdisciplinarity. We define the \revisiontwo{list} 
  $\bm{SC}_i$ of \textit{social codes or categories} of author $i$ as
  the union of the \revisiontwo{lists} of personal PACS codes or research categories
  of all the co-authors of $i$, excluding the codes that are already
  in $PC_i$. We then measure the \textit{social interdisciplinarity}
of author $i$ through the \textit{social entropy} $S_i$ defined as the
Shannon entropy of the \revisiontwo{list} $\bm{SC}_i$:
\begin{equation}
S_i = - \sum_{\alpha} q\lay{\alpha}_i \log (q\lay{\alpha}_i),
\label{eq:exposure_entropy}
\end{equation}
where $q\lay{\alpha}_i$ is the frequency of code or category $\alpha$ in $\bm{SC}_i$. 

\subsection{Interdisciplinarity and success} 
{\color{black}

For different values of background entropy $B$,
Fig.~\ref{fig:fig1}(a,b) shows the average cumulative number of
citations $\citsymbol$ received by comparable authors with career
lengths ranging between $5$ and $15$ years and with a number of
publications ranging from $5$ to $100$. The vignettes in
Fig.~\ref{fig:fig1}(a,b) illustrate the typical compositions of the
two \revisiontwo{lists}  $\bm{PC}_i$ and $\bm{PC}_j$, respectively for a specialized
author $i$ and for an interdisciplinary author $j$. On average,
authors with intermediate values of background entropy (i.e., neither
interdisciplinary nor specialized) are characterized by a relatively
low value of scientific performance. Both in physics
(Fig.~\ref{fig:fig1}(a)) and in the natural sciences
(Fig.~\ref{fig:fig1}(b)), scientific performance exhibits a U-shaped
trend, with a minimum around $B_i\simeq 0.2$ and a maximum at
$B_i\simeq 1.6$ for APS and at $B_i\simeq 1.4$ for WOS. Interestingly,
a similar and relatively large number of citations can be obtained
equally by highly specialized and highly interdisciplinary
authors. However, the asymmetry in the U-shaped trend indicates that,
on average, scientists with the widest range of research interests
($B_i \gtrsim 1.4$) tend to outperform not only less interdisciplinary
scientists, but also the most specialized ones ($B_i\simeq 0$).}
Overall, these results suggest that, {\color{black} at the micro level
  of physics as well as at the macro level of the natural sciences,}
both specialized and interdisciplinary scientists can be successful;
yet extreme interdisciplinarity provides competitive advantage over
extreme specialization. Moreover, success is thwarted as scientists
abandon extreme specialization in favor of moderate degrees of
interdisciplinarity.

\revisiontwo{Next, we examine whether scientists' performance is associated not only with their background, but also with the variety of opportunities
they can tap through their collaborators} \cite{Sayama}. {\color{black}
  For the same subset of authors as in Fig.~\ref{fig:fig1}(a,b),
  Fig.~\ref{fig:fig1}(c,d) shows the relationship between average
  number of citations and social entropy.} Results indicate that
authors can amplify success as their social interdisciplinarity
increases. An author with a more heterogeneous network (i.e., a higher
value of social entropy $S$) will, on average, have a higher
performance than an author with a more homogeneous network (and lower
$S$). Thus, while specialization can be a successful strategy
(Fig.~\ref{fig:fig1}(a,b)), seeking collaborators with few and
overlapping specialities will be a hindrance. Scientists can instead
enhance their performance by selecting interdisciplinary collaborators. 
 
  To test whether our findings differ from what would be
  expected if codes or research categories were randomly assigned to
  articles, we replicated the analysis in Fig.~\ref{fig:fig1} using a
  null model that preserves the co-authorship network and the number
  of citations and codes per article, but in which codes and
  categories are randomly reshuffled across articles published within
  the same year.
Fig. 2 shows the observed relationship between background and social
interdisciplinarity for subsets of authors with a different number of
citations accrued, compared with the relationship found
  in the corresponding null model. Interestingly, the more successful
the authors, the larger and more correlated the values of their
background and social interdisciplinarity. These results provide
evidence of an interplay between scientists' personal interests and
expertise, on the one hand, and their collaborative patterns on the
other, and suggest that such interplay is associated with scientific
performance.

\begin{figure}[ht]
\begin{center}
\includegraphics[width=3.in]{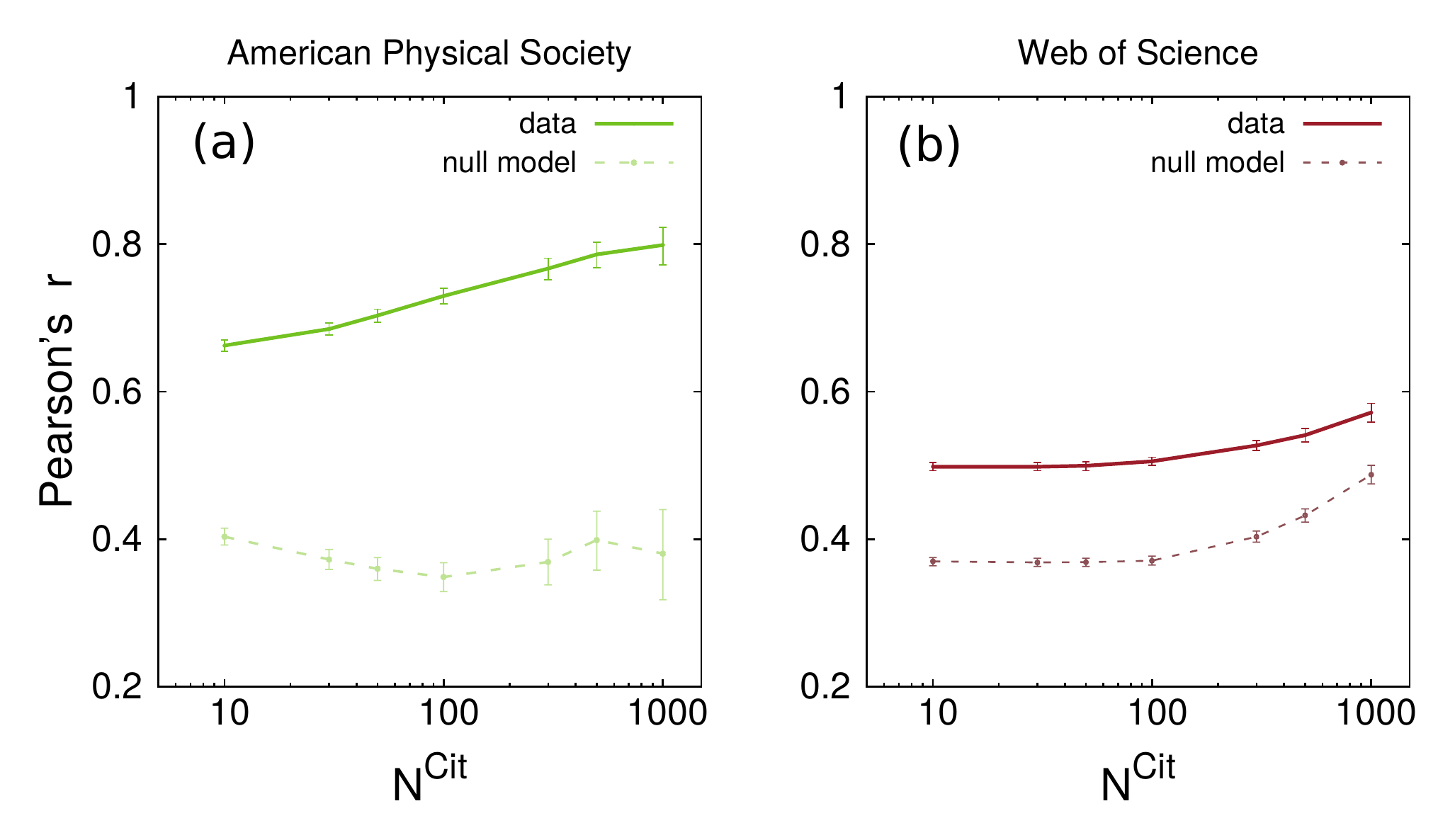}
\end{center}
\caption{\textbf{Interplay between background and social interdisciplinarity.} The relationship between the two forms of interdisciplinarity plays a crucial role in sustaining success: the values of background and social entropy become more correlated as authors are associated with more citations, both in physics (panel a) and in the natural sciences (panel b). The Pearson linear correlation coefficient $r$ between the two quantities consistently increases from $r\simeq 0.65$ (APS) and $r\simeq 0.5$ (WOS) to $r\simeq 0.8$ (APS) and $r\simeq 0.6$ (WOS) as the number of authors' citations increases from values smaller than $100$ to values around $1,000$. The shaded circles along the dashed lines refer to values of the correlation coefficient obtained from a null model in which PACS codes and categories are randomly reshuffled across articles, while the number of citations, articles, codes and categories per author is preserved. Error bars represent $95$\% confidence intervals based on Fisher's $r$-to-$z$ transformation.
}
\label{fig:fig2}
\end{figure}

So far we have combined authors with different career lengths and productivity, and from different research fields. We shall now examine the relationship between interdisciplinarity and success by accounting for potential biases that may arise from confounding factors such as heterogeneity in the length of scientists' careers and in the patterns of citations characterizing different periods and research fields.

\subsection{Comparing different career stages}

\begin{figure*}[ht]
   \centering
   \includegraphics[width=5.20in]{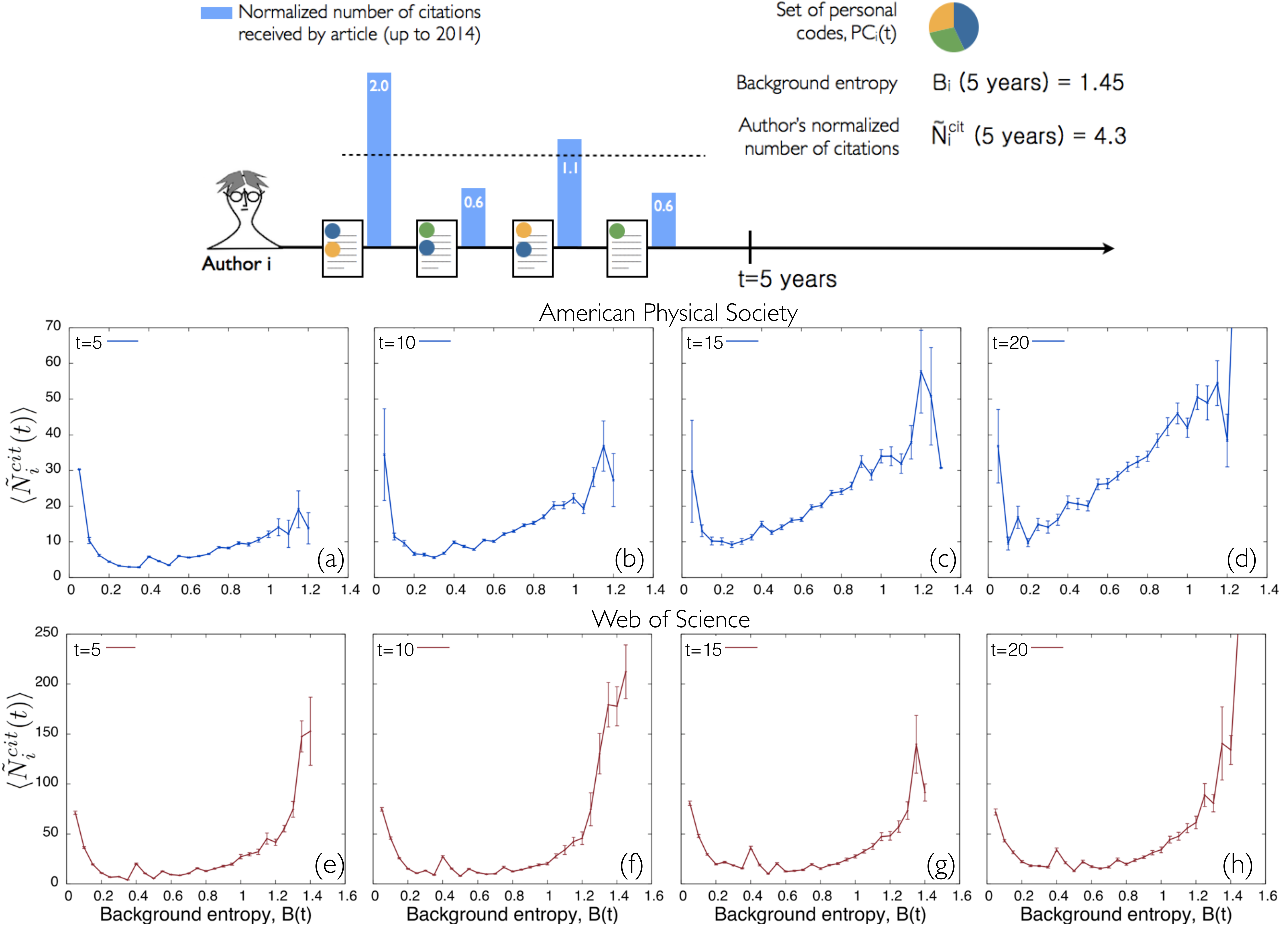}
   \caption{\textbf{Interdisciplinarity and success at different career stages.} {\color{black} The normalized number of citations obtained by authors at different career stages as a function of their background entropy in physics (panels a-d) and the natural sciences (panels e-h). The vignette illustrates how performance and interdisciplinarity were measured. For each author $i$ and a given year $t$ of $i$'s career, both performance and interdisciplinarity were measured at $t$ on all articles published by $i$ since the beginning of $i$'s career up to $t$. The citations accrued by each article up to $t$ were normalized through the method proposed in \cite{Radicchi2011}. The U-shaped dependency of $\langle \tilde{N}_{i}^{cit}(t) \rangle$ on background entropy and the presence of a minimum at intermediate values of $B(t)$ characterize both young authors and experienced ones, thus indicating that extreme interdisciplinarity as well extreme specialization are already beneficial at the very beginning of a scientist's career. However, the competitive advantages of background interdisciplinarity become more pronounced as careers progress toward their final stages when the difference in performance between the most interdisciplinary and the most specialized authors reaches its peak. Error bars represent the standard error of the mean.
} 
}    
\label{fig:fig3}
\end{figure*}

For each year $t$ of scientist $i$'s career, we consider the
background entropy $B_{i}(t)$ and the social entropy $S_{i}(t)$
calculated, respectively, on the set of $i$'s publications and
collaborations up to $t$. Moreover, drawing on the method proposed in
  \cite{Radicchi2011}, we measure the performance of scientist $i$ at
  year $t$ through the {\em normalized} number of citations
  $\tilde{N}_{i}^{cit}(t)$. This enables us to account for variations in: (i) patterns and
  volume of citations across sub-fields and disciplines; (ii)
  attractiveness of research topics over time; and (iii) the starting
  year and duration of authors' careers. To obtain
  $\tilde{N}_{i}^{cit}(t)$, we compute the normalized number of
  citations of a given article $a$ as the ratio between the total
  number of citations received by article $a$ and the average number
  of citations received by all articles published in the same
  sub-field or discipline and year as $a$ \cite{Radicchi2008}. We then
  sum the normalized numbers of citations of all articles published by
  author $i$ in each year up to $t$, and obtain
  $\tilde{N}_{i}^{cit}(t)$ \cite{Radicchi2011} (see vignette in
  Fig.~\ref{fig:fig3}). For instance, by evaluating $B_i(5)$, $S_i(5)$
  and $\tilde{N}_{i}^{cit}(5)$, we can compare values of background
  and social interdisciplinarity, and normalized number of citations,
  respectively, at the end of the first five years of each author
  $i$'s career. Following \cite{Radicchi2011}, we assess interdisciplinarity through the
  second level (i.e., the first four digits) of the PACS hierarchy, at
  which the universality of scaling holds.

Fig.~\ref{fig:fig3} reports the normalized number of
  citations $\langle \tilde{N}_{i}^{cit}(t) \rangle$ averaged over
  authors characterized by a certain value of background
  interdisciplinarity at four career stages, namely at $t=5,10,15,20$
  years, in physics (panels a-d) and in the natural sciences (panels
  e-h). Interestingly, the U-shaped functional form of the
  relationship between success and background interdisciplinarity is
  found already at the fifth year of a scientist's career, and
  persists across all career stages, both in physics and at the
  broader level of the natural sciences. In particular, scientists
  with a value of background entropy $0.2 \lesssim B_i \lesssim 0.6$
  have a poorer performance than scientists that are either highly
  specialized or highly interdisciplinary. Similarly, social
  interdisciplinarity is associated with scientific success
  across all stages of a scientist's career.

Fig.~\ref{fig:fig3} suggests that the relationship between
interdisciplinarity and success is subject to a temporal drift. At the macro level of the natural
  sciences, while the maximum normalized number of citations accrued
  by the most interdisciplinary author $i$ at the fifth year of career
  is, on average, $\tilde{N}_{i}^{cit}(5) \simeq 160$, the largest
  value of $\tilde{N}_{j}^{cit}(20)$ for an author $j$ at the 20-th
  year of career is, on average, $\tilde{N}_{j}^{cit}(20) \simeq
  250$. Thus, an author $i$ with, for instance, $B_i \simeq 1.0$ at
  the fifth year of $i$ 's career will have, on average, $30$
  normalized citations, namely $0.40$ times as many citations as those
  accrued by the most specialized author $j$ (i.e., with $B_j\simeq
  0$), and only $0.20$ times as many citations as those of the
  best-performing author (i.e., with $B_j\simeq 1.40$). At the 20-th
  year of his or her career, the same author $i$ with $B_i\simeq 1.0$
  would still be able to accrue, on average, about $30$ normalized
  citations. However, while $i$'s comparative disadvantage over the
  most specialized author would remain unaltered, the disadvantage
  over the most interdisciplinary one would further
  intensify. Author $i$ would therefore need to keep increasing
  background interdisciplinarity over time, lest by the 20-th year of
  $i$'s career the total number of citations be, on average, only
  $0.12$ times as large as the one of the best-performing author. \revisiontwo{The
association of background interdisciplinarity with success thus becomes stronger as scientists' careers progress.} Moreover, in the long run,
as careers approach their final stages, not only are highly
interdisciplinary scientists more successful than specialized ones,
but the difference in performance between the most
  interdisciplinary and the most specialized scientists reaches its
  peak.

\subsection{Paths to interdisciplinarity}

\begin{figure*}[ht]
   \centering
   \includegraphics[width=6in]{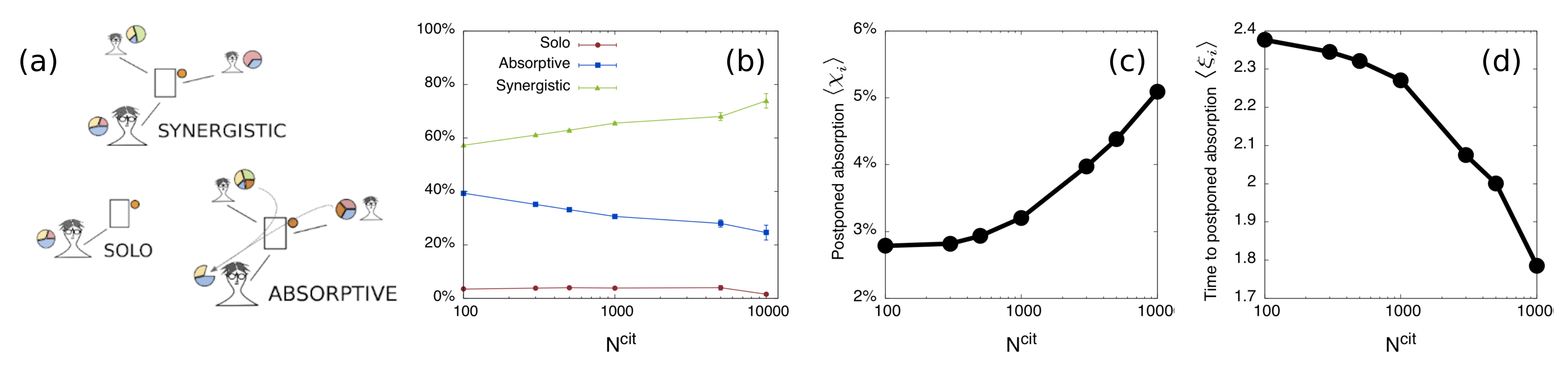}
   \caption{\textbf{Strategies for enhancing interdisciplinarity.} 
{\color{black} (a) We identify three main strategies through which
       authors can expand their knowledge into a new field: (i) by
       publishing on their own in the new field (solo strategy);
       (ii) by collaborating with others that have already published
       in the field (absorptive strategy); and (iii) by
       collaborating with others that have never published in the field (synergistic strategy). (b)  For authors in the APS data set  whose articles obtained more than a given number of citations, we measured the average frequencies of solo, absorptive and synergistic strategies. Successful authors are more prone to synergistic strategies than less successful ones. Error bars represent the standard error of the mean. (c) The average fraction of social PACS codes eventually acquired by an author is positively correlated with the author's success. (d) The average time needed to acquire new PACS codes from collaborators is negatively correlated with an author's success. Successful authors are more likely not only to acquire knowledge from their collaboration network, but also to do so more quickly than less successful ones.
}}
\label{fig:fig4}
\end{figure*}

Given the advantages of interdisciplinarity, how do scientists widen
their background over time, and which research strategies are
associated with success? We identify three strategies: solo,
absorptive and synergistic. First, we define the \textit{solo strategy} as the 
  acquisition of new knowledge through the publication of a
single-authored article in the corresponding scientific area. With
this strategy, scientists extend their background
  interdisciplinarity through ``in-breadth'' learning; yet they do
not amplify their social exposure to new sources of knowledge. Second,
the \textit{absorptive strategy} is defined as the 
  acquisition of new knowledge by an author through the publication
of a multi-authored article with at least one co-author who has
already published in the corresponding area.  Through
  this strategy, scientists absorb knowledge from their collaborators
  as soon as they are exposed to it, thus increasing their background
interdisciplinarity (and possibly the heterogeneity of their
collaboration networks). Finally, the \textit{synergistic strategy} is
defined as the acquisition of new knowledge by an
author through the publication of an article with co-authors who have
never published in the corresponding area. Through this strategy,
collaboration promotes cross-fertilization of various disciplinary
areas, and ultimately intensifies all co-authors' (background)
interdisciplinarity through the acquisition of new knowledge. An
illustration of the three strategies is reported in
Fig.~\ref{fig:fig4}(a).

We denote by $P_{i}^{\rm solo}$, $P_{i}^{\rm abs}$ and $P_{i}^{\rm syn}$ {\color{black} the fraction of number of PACS codes} acquired by author $i$ through, respectively, the solo, absorptive, and synergistic strategies, during $i$'s entire career. To understand how authors with different performance vary in their usage of the three strategies, in Fig.~\ref{fig:fig4}(b) we show the average frequencies of solo, absorptive, and synergistic strategies {\color{black} adopted by authors in the APS data set whose articles accrued a total number of citations exceeding various thresholds. Remarkably, the overall frequency of the solo strategy is just about $4\%$ at all levels of success, whilst the vast majority of the new PACS codes (about $96\%$) originate from collaboration. In particular, not only are authors across all levels of performance more likely to embrace a new sub-field through a synergistic strategy than an absorptive one, but also the difference in usage frequencies between the two strategies widens as authors are more successful ($N^{cit} \sim 10,000$).}  

{\color{black} Exposure to collaborators' knowledge may broaden a scientist's background interdisciplinarity not only instantaneously through the absorptive strategy.} When engaged in a joint endeavor, scientists can, in principle, gain access to the entire spectrum of knowledge offered by their collaborators. {\color{black} A fraction of this knowledge can indeed be absorbed as soon as collaboration occurs, through the absorptive strategy (Fig.~\ref{fig:fig4}a); the remaining can be acquired at a subsequent stage,} through a process here referred to as \textit{postponed absorption}. Diluting acquisition of new knowledge over time can have various effects on performance, depending on how much knowledge is acquired and the time separating acquisition from exposure.

{\color{black} To study the degree to which knowledge acquisition is
  affected by past collaborations, for each author $i$ we quantify the
  propensity, once exposed to a new social PACS code $\alpha$,} to
acquire it at some subsequent stage. Exposure to sub-field $\alpha$
occurs when author $i$, with no experience in $\alpha$, for the first
time collaborates with someone who has already published in
$\alpha$. {\color{black} Postponed absorption of $\alpha$ occurs when,
  for the first time after exposure, $i$ appears as the solo author or
  co-author of an article $a$ in $\alpha$. Notice that the co-authors
  of $a$ are assumed not to have experience in $\alpha$ (or else
  knowledge acquisition would be classified as instantaneous
  absorption). Of the social PACS codes to which author $i$ was
  exposed, we measure the fraction $\chi_i$ that was eventually
  acquired by $i$.} Lastly, for each author $i$ we measure the mean
interval of time $\xi_i$ separating postponed absorption from exposure.

Fig.~\ref{fig:fig4}(c) shows the average fraction $\avg{\chi_i}$ over all authors, and suggests that successful authors in the APS data set are more likely to use up their collaboration networks to acquire new knowledge than less successful ones. {\color{black}Moreover, Fig.~\ref{fig:fig4}(d) shows the average interval of time $\avg{\xi_i}$ over all authors, and suggests that the time separating exposure to new knowledge from acquisition tends to become shorter as authors' performance increases. Not only do successful scientists choose their collaborators carefully so as to secure exposure to new areas, but they also prefer not to wait too long before they publish in those areas either on their own (solo strategy) or with other collaborators (synergistic strategy).} 

\section{Discussion}

This study was concerned with intellectual diversity and diversification in science. First, we showed that there are larger returns to ``in-breadth'' than ``in-depth'' learning, especially as scientists' careers progress. However, scientists bear opportunity costs as they begin to diversify their background, at least until they become highly interdisciplinary. Second, we found that scientists benefit from heterogeneous collaboration networks. Scientists with groups of collaborators spanning many different areas are more successful than those with collaborators focused on one or few overlapping areas. Third, results indicated that successful scientists tend to match the diversity of their background with the diversity of their collaborators. 

\revisiontwo{Recent work on interdisciplinarity has focused mainly on the benefits and penalties associated with authors' diversity of knowledge and background \cite{Leahley,Chakraborty,Sayama}. However, research on social capital and innovation has also suggested that performance is nurtured by the opportunities of knowledge recombination offered by the network in which individuals are embedded ~\cite{Singh,Burt,Rodan,Scholtes2014,Sosa2011a}. In our study, we integrated the individual and social perspectives, and proposed a conception of interdisciplinarity} that extends beyond the boundaries of the scientists' background to also include their collaboration networks. We suggested that scientists can integrate and extend ``in-breadth'' learning by widening the breadth of their social network. This idea of trading off learning against collaboration is akin to models of embodied, environmentally embedded cognition put forward by recent developments in cognitive science and philosophy of mind \cite{Clark2008}. Individuals can amplify their cognitive abilities and processes by exploiting external resources in their physical and social environment. Similarly, we have shown that scientists can enhance success by retrieving and recombining external pools of knowledge.

\revisiontwo{Previous work has documented an increase in multi-authored articles over the last few decades \cite{Wuchty2007}, and has suggested that successful authors tend to develop interdisciplinary careers by specializing on various distinct topics at any particular time \cite{Chakraborty}. We integrated and extended these studies by investigating the collaborative strategies that sustain knowledge diffusion and acquisition}. Intellectual diversification is most effective when it is pursued through absorption of new knowledge from others or through joint endeavors aimed at producing new knowledge. Lone scientists toiling away at extending their competence into new research areas tend to be less successful than those who instead leverage on social networking to extract new knowledge from collaborators \cite{Singh}. Lacking visibility and credibility can also pose barriers to entry into hitherto uncharted territories of science \cite{Sinatra2015,PetersenInequality2014}. Relying on collaborators' experience can help reduce such barriers and facilitate access to multiple scientific communities.

Our analysis is not without limitations, {\color{black} chiefly concerned with the use of citation-based metrics as indicators of
scientific merit \cite{Bormann,Penner2013predictability,Petersen2014,Radicchi2008} and with the generalizability of results beyond the natural sciences.} Despite this, our study has far-reaching implications for research and policy. Opening up the black box of the scientist's knowledge to also account for collaboration networks paves the way for more integrated approaches to scientific production that borrow insights from bibliometrics and citation analysis, complex networks, cognitive science and the sociology of science. Our findings can also inspire individual scientists to shape and sustain successful careers, research institutions to strengthen their scientific reputation and profile through effective recruitment policies and internal evaluation systems, and funding bodies to award research grants to projects with the highest potential impact.


\end{document}